\documentclass[aps,prl,twocolumn,superscriptaddress,nofootinbib]{revtex4}
\usepackage{graphicx}
\begin{document}

\title{Method for Computing Short-Range Forces between Solid-Liquid
Interfaces driving Grain Boundary Premelting}

\author{J. J. Hoyt$^1$, David Olmsted$^2$, Saryu Jindal$^3$, Mark Asta$^3$,
Alain Karma$^4$}
\affiliation{$^1$Department of Materials Science and Engineering, McMaster
University,\\
Hamilton, ON, Canada\\
$^2$Sandia National Laboratories, Albuquerque, NM\\
$^3$Department of Chemical Engineering and Materials Science,
University of California, Davis, CA\\
$^4$Department of Physics and Center for Interdisciplinary Research on
Complex Systems,
Northeastern University, Boston, MA}

\begin{abstract}
We present a molecular dynamics based method for computing accurately
short-range structural forces resulting from the overlap of spatially diffuse 
solid-liquid interfaces at wetted grain boundaries close to the
melting point. The method is based on monitoring the fluctuations of the
liquid layer width at different temperatures to extract  the excess
interfacial free-energy as a function of this width.  The method is
illustrated for a high energy $\Sigma$9 twist boundary in pure Ni. The
short-range repulsion driving premelting is found to be dominant in
comparison to long-range dispersion and entropic forces and consistent with
previous experimental findings that nanometer-scale layer widths may only
be observed very close to the melting point.
\end{abstract}
\vspace{1cm}

\maketitle

The term premelting refers to the formation of a thin, thermodynamically
stable, liquid-like film at an interface for temperatures below the equilibrium
melting point ($T_M$).  Premelting at grain boundaries (GB) can have dramatic 
consequences in the context of materials processing, and the physical 
properties of polycrystals at high homologous temperatures.  Despite the 
importance of this phenomenon, direct experimental observations of GB 
premelting remain relatively rare, \cite{expt_list2,expt_list1,expt_list3} 
particularly in the case of pure materials \cite{expt_list1,expt_list3}.  
Consequently, outstanding fundamental questions remain concerning the nature 
of the forces which drive premelting at these internal interfaces.  In this 
Letter we introduce a molecular dynamics (MD) method that exploits large 
fluctuations in GB width to compute short-range forces resulting from the 
overlap of spatially diffuse crystal-melt interfaces from two grains of 
different orientations.  We demonstrate the application of this method in a 
direct calculation of the excess free-energy of the GB as a function of this 
width for a high-energy boundary in a classical model of elemental Ni.  The 
results yield quantitative insights into the relative magnitudes of these 
short-range structural forces and other long-ranged contributions, and help
explain the origins of the experimental observations \cite{expt_list3} that
GB premelting in pure metals may occur only over extremely small temperature 
ranges near $T_M$.

Premelting generally reflects a competition between opposing bulk and 
interfacial thermodynamic factors, giving rise to a free energy (per unit 
area) of the following form (e.g., \cite{Lipowsky}):
\begin{equation}
G(w)=\Delta G_f w + \Psi(w).
\label{eq1}
\end{equation}
In Eq.~\ref{eq1} $w$ represents the width of the premelted layer, $\Delta G_f$ 
is the free energy difference between solid and liquid (per unit volume) that 
penalizes the formation of liquid films below $T_M$, and $\Psi(w)$ is the 
so-called ``disjoining potential" which takes the limits of $\gamma_{GB}$ (the 
interfacial free energy of a ``dry" grain boundary) and $2 \gamma_{SL}$ (twice 
the solid-liquid interfacial free energy) for zero and infinite $w$, 
respectively.  In general, the disjoining potential contains both repulsive and 
attractive contributions.  Long-ranged dispersion forces lead to an attractive 
interaction between solid-liquid interfaces \cite{Lipowsky,Clarke} which are 
dominant at large $w$ and are predicted to give rise to finite interfacial 
widths at $T_M$ \cite{Lipowsky}.  For 
$\Delta \gamma =  \gamma_{GB} - 2 \gamma_{SL} > 0$, a repulsive contribution to
$\Psi(r)$ arises from short-ranged structural interactions ($\Psi_{sr}$), 
associated with the overlap of the diffuse regions of the solid-liquid 
interfaces.  The exact nature of this structural contribution remains less well 
understood.

Mean-field arguments \cite{Widom}, as well as lattice-gas models (e.g., 
\cite{Kikuchi}), yield an exponentially decaying form for the short-ranged 
contribution to the disjoining potential:
\begin{equation}
\Psi_{sr}(w)=2 \gamma_{sl} + \Delta \gamma \exp [-w/\delta]
\label{eq3})
\end{equation}
where $\delta$ is an interaction length on the order of the atomic spacing.  In the 
absence of long-ranged dispersion forces, and neglecting capillary fluctuations which 
lead to an additional repulsive contribution to $\Psi(r)$ (see below), insertion of 
Eq.~\ref{eq3} into Eq.~\ref{eq1} leads to the prediction of a continuous premelting 
transition with an equilibrium grain boundary width that diverges logarithmically as 
$T_M$ is approached from below.  Recent theoretical results suggest that the nature 
of $\Psi_{sr}$ can be much more complex.  Diffuse-interface theories, 
\cite{Lobkovsky,Carter} which neglect long-ranged forces and capillary fluctuations, 
have shown that the dependence of $w$ on temperature may in some cases display a 
discontinous jump, with the coexistence of ``wet" and ``dry" interface states, 
while other parameter choices lead to continuous increases in $w$ up to $T_M$.  In 
recent applications of the phase-field crystal (PFC) method to the study of grain 
boundary premelting \cite{Berry,Mellenthin}, results for 2D hexagonal systems
give a disjoining potential that is purely repulsive above
a critical misorientation \cite{Mellenthin}, but exhibits a minimum, corresponding
to a finite layer width at the melting point, below it.
While these 
theoretical models thus suggest a rich behavior for the disjoining potential 
in general, it remains unclear for which GB misorientations the various qualitative
forms for $\Psi_{sr}(w)$ may be expected in
real materials, and how structural short-range forces compare qualitatively to 
long range forces.
To facilitate further progress in the understanding of the forces
that drive premelting, we describe in the remainder of this Letter a quantitative 
framework for the direct calculation of $\Psi_{sr}(w)$, through histogram analyses 
of interface widths derived from MD.

Classical MD simulations provide a framework ideally suited for probing the 
short-ranged structural contributions to $\Psi(w)$.  Such simulations have been 
employed extensively in the past to study GB premelting \cite{md_list,sutton} 
and in the present work we propose a methodology to extend the analysis of such 
MD results as a framework for extracting $\Psi(w)$.  We demonstrate the approach for
a classical model of elemental Ni, described by the embedded-atom potential of
Foiles, Baskes and Daw \cite{FBD}.  The potential was chosen as we have previously
calculated the solid-liquid interfacial free energies, melting temperature
and solid-liquid thermodynamic properties with high precision.  A value for 
$\gamma_{SL}$ of $285$ $mJ/m^2$ for the potential has been determined using the 
capillary fluctuation method \cite{Hoyt2}, and a coexistence technique was used to 
compute a melting temperature of $1710\pm5K$ \cite{Sun} (from subsequent coexistence 
runs and an analysis of GB width fluctuation data presented below the uncertainty in 
this estimate has been reduced to approximately 1 degree).  We began by considering a 
total of four boundaries with a range of zero-temperature grain-boundary energies 
spanning 450 to i143i mJ/m$^2$, which is 0.9 to 2.5 times the value of $\gamma_{SL}$.
For each GB we performed a conjugate gradient minimization (exploring also the 
microscopic translational degrees of freedom and the excess number of atoms at the
grain boundary) to derive an optimized zero-temperature interface structure.  With 
this structure as a starting point the GBs were heated gradually up to the melting 
point employing constant-temperature MD simulations following the procedure described 
in the on-line supplemental material.

\begin{figure}
\includegraphics[scale=.8]{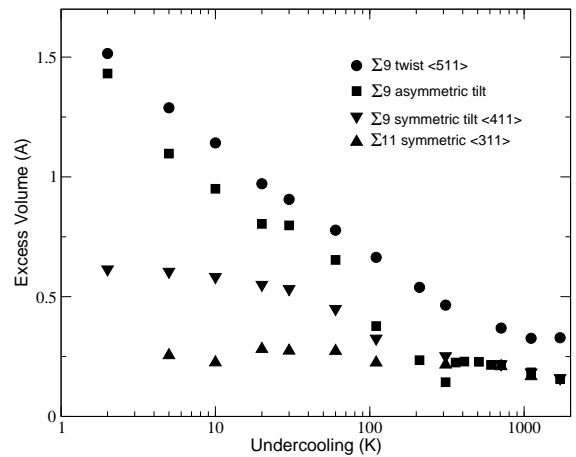}
\caption{\label{excess} Calculated excess volumes versus temperature for four grain
boundaries in elemental Ni.}
\end{figure}

Figure~\ref{excess} shows the calculated excess volume of each GB, displaying three
qualitatively different behaviors.  The highest energy boundaries feature an excess 
volume displaying a logarithmic divergence characteristic of a continuous premelting 
transition.  The lower energy GBs show two different behaviors; in one case the 
excess volume rises with increaing temperature but then plateaus, maintaining a 
finite excess volume at the melting temperature.   The lowest energy GB shows an 
excess volume that is relatively small and only weakly dependent on temperature.  The 
range of behavior demonstrated by the different GBs in Fig.~\ref{excess} is 
qualitatively similar to that seen in very recent GB simulations for Si \cite{sutton}.

For the remainder of this Letter we focus on one of the two high-energy
boundaries displaying clear
premelting behavior, namely the $\Sigma$9 boundary characterized by a
$120^o$ rotation about
the GB normal lying along the $[511]$ crystallographic direction.  We study
the structural
properties of the grain boundary at five temperatures over a range of 
undercoolings 30 to 2 K
below $T_M$.  After
4.2 ns equlibration runs at each temperature, statistics were obtained for
$w$ at a given
temperature as follows.  For each snapshot (selected at a frequency of 10
ps) $w$ is determined
by utilizing the scheme developed in the capillary fluctuation method \cite
{Hoyt2}.  Each atom
is assigned a structural order parameter, $\phi_i$, constructed from the
positions of
the 12 nearest neighbor atoms and
the $\phi_i$ values are then averaged in bins along the
direction normal to the boundary.  The point of inflection in the average
order parameter
profile is taken as the position of one of the solid-liquid interfaces.
As described in more detail in the accompanying on-line material, the
procedure is
repeated to locate the second solid-liquid interface and hence the GB
width.

\begin{figure}
\includegraphics[scale=.4]{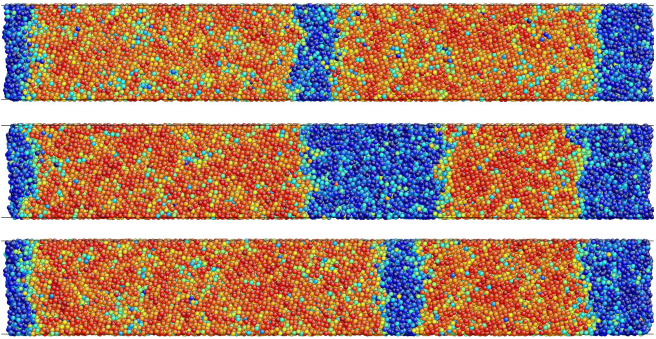}
\caption{\label{snapshots}Snapshots from an MD simulation at an undercooling 
of 2 K illustrating the dynamic nature of the GB width.  The red (blue) 
areas indicate regions of solid (liquid) like liquid order.  The liquid-like
retions at the far left and right hand sides represent premelting of the free 
surfaces of the simulation cell, while that in the middle corresponds to the
premelted grain boundary.}
\end{figure}

\begin{figure}
\includegraphics[scale=.3]{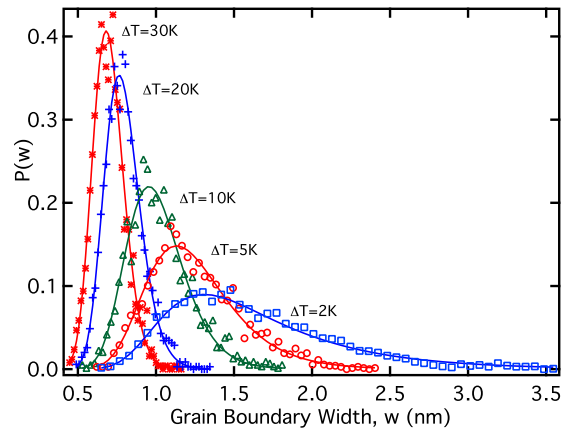}
\caption{\label{distr}The distribution function $P(w)$ vs $w$ from the MD
simulations (symbols)
versus the least square fits of the premelting model of Eqs.~\ref{eq1} and
\ref{eq3}.}
\end{figure}

An important observation in the present work is that
the width of the GB regions is highly dynamic in the MD
simulations, particularly
at the temperatures closest to $T_M$.  This point is illustrated clearly in
Fig.~\ref{snapshots}
which shows three snapshots, taken from a 40 ns simulation at an
undercooling of 2 K, where the
atoms have been color coded based on their $\phi_i$ values; blue
representing a liquid-like
environment and red the crystal.
The snapshots clearly
demonstrate the presence of large fluctuations in the width of the
premelted layer over
the course of the simulation.
The highly dynamic nature
of the premelted layer provides a framework for extracting the disjoining
potential.
We show in Fig.~\ref{distr} histograms of interface width obtained
at five temperatures near the melting point.  The solid lines represent
least-squares fits to the
data employing the thermodynamic model of Eq.~\ref{eq1} as follows.  The
probability ($P(w)$) of
observing a premelted layer  width $w$ is given as:
\begin{equation}
P(w)=C \exp [-A G(w)/k_B T]
\label{eq2}
\end{equation}
where $C$ is a temperature-dependent normalization constant, $A$ is the
cross-sectional area
and $G(w)$ is defined in Eq.~\ref{eq1}.  The data in Fig.~\ref{excess}
suggests a logarithmic
divergence of $w$ with increasing temperature, and in order to fit Eq.~\ref
{eq2} to the MD
data, we therefore employ the form for the disjoining potential given in
Eq.~\ref{eq3}.  The
least square fits of $P(w)$ versus $w$ for all five undercoolings studied
are shown in
Fig.~\ref{distr}.  The excellent agreement suggests that the free energy of
Eq.~\ref{eq1}
together with the disjoining potential (Eq.~\ref{eq3}) represents an
accurate model for the
premelting behavior of this boundary.  The analysis  of Fig.~\ref{distr}
assumed a melting
point of $T_M=1710 K$.  If instead a value of just one degree different,
i.e., $T_M =1709$,
is assumed, then a poor fit to $P(w)$ is obtained at the lowest
undercooling.  In addition,
for simulations run at a temperature of $1712 K$ the system exhibited a
gradual melting.
These findings, together with the results of separate coexistence
simulations, indicate that
the melting temperature is known to a precision approaching $\pm 1^o$.

As an independent check on the validity of Eq.~\ref{eq3}, we employ an 
additional analysis of the data of Fig.~\ref{distr}. From Eq.~\ref{eq2} the 
disjoining potential can be written in terms of $P(w)$ as:
\begin{equation}
\Psi(w) = -(k_B T/A) ln P(w,T_i) - \Delta G_f w + a_i
\label{eq4}
\end{equation}
where the $a_i$ are unknown constants related to $C$ in Eq.~\ref{eq2} and the
subscript $i$ denotes a separate histogram of data corresponding to each 
undercooling.  The $a_i$ can be determined by a least square fitting procedure 
such that all the data sets can be merged and the entire function $\Psi(w)$ 
constructed.  Notice the procedure adopted here is analogous to the histogram
method, often employed in Monte Carlo simulations to extract transition states 
and energy barriers, but with the undercooling playing the role of a bias 
potential.  The results of the histogram procedure are shown in Fig.~\ref{distr}. 
The inset of the figure plots the right-hand side of Eq.~\ref{eq4}, with all the 
constant terms $a_i$ set to zero to illustrate that different undercoolings 
sample a range of $w$ regimes of $\Psi(w)$.  The main figure shows the final 
$\Psi (w) $ function along with a fit (solid line) to the exponential form given
in Eq.~\ref{eq3}.  It is important to note that the fit parameters obtained
via the histogram method ($\delta= 2.49 \AA$ and $\Delta \gamma = 156mJ/m^2$)
compare very well to those derived through the individual fits of the separate 
histograms in Fig.~\ref{distr}:
$\delta = 2.67 \pm 0.18 \AA$ and $\Delta \gamma = 127 \pm 26 mJ/m^2$.  

\begin{figure}
\includegraphics[scale=.3]{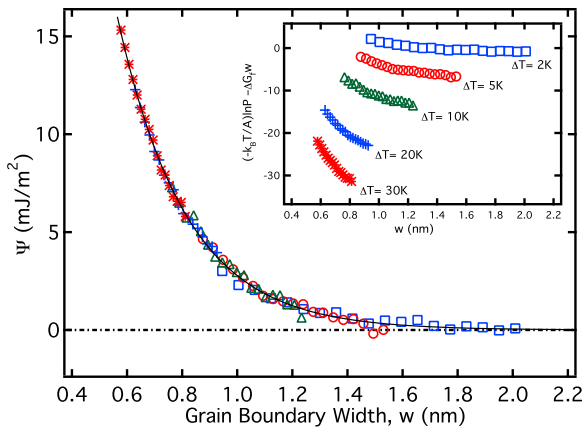}
\caption{\label{histfig}An illustration of the histogram method used to
extract the disjoining potential.  The inset plots the right hand side of 
Eq.~\ref{eq4} with the constants $a_i$ set to zero.  The main plot merges the 
individual histograms of data and reproduces the complete disjoining 
potential $\Psi (w)$.  The solid line is a best fit to the exponential decay 
given in Eq.~\ref{eq3}.}
\end{figure}

The present simulations have quantified only the short-ranged contribution
to $\Psi(w)$.  However, for setting the temperature scale over which premelting 
can be observed in elemental metals, we estimate that this is the dominant 
contribution.  For values of $w \sim$ 1 nm we can estimate an upper bound on the 
dispersion forces using a value of the Hamaker constant measured for surface 
premelting of metals \cite{Pluis}.  This gives a contribution to
$\Psi$(w=1 nm) $\sim .4$ mJ/m$^2$, approximately one order of magnitude
smaller than the results presented in Fig. 4, computed from the value of 
$\Psi_{sr}$ using the parameters derived above.  There is also a known
entropic contribution to $\Psi(w)$ associated with the long-wavelength
fluctuations of the two interfaces for large separation \cite{LipowskyFisher}. 
This contribution is generally repulsive because the conditions that the two 
interfaces do not intersect reduces the available configurational entropy of 
interface meandering. While this entropy reduction produces a physically important 
long-range force for one-dimensional interfaces, such as step ledges on surfaces, 
it produces only a subdominant short-range forces for two-dimensional interfaces 
owing to the slow logarithmic growth of the mean-square fluctuation amplitude with 
interface area, as compared to the much faster square-root growth of this amplitude 
with interface length in one dimension.  In particular, a straightforward estimate 
of this force using analytical results of the literature \cite{LipowskyFisher} and
the parameters for Ni from the MD results show that this entropic force is
completely negligible in comparison to the short-range structural forces
computed here.  

Thus, for high-energy boundaries it can be expected that the lowest temperatures 
where premelting will become appreciable is set through the relation
$(\Delta T / T_M) = (\Delta \gamma / L \rho \delta) \exp{(-w_{eq}/\delta)}$, with 
$w_{eq} \sim$ 1 nm, where we have expressed $\Delta G_f = L \Delta T \rho / T_M$ 
in terms of the latent heat per atom ($L$), the solid density ($\rho$) and the 
undercooling ($\Delta T = T_M - T$).  The present results give $\delta$ on the 
order of an interatomic spacing and $\Delta \gamma$ on the order of half
$\gamma_{SL}$.  While the exact values will vary somewhat depending on the system, 
we believe these values are well representative of high-energy boundaries in pure 
metals.  With these estimates we obtain a value of $\Delta T$ required to obtain 
$w \sim 1 nm$ of $(\Delta T / T_M) \sim (\alpha/2) \exp{(-4)}$, where we have used
$\rho \delta^{1/3} \sim 1$ and the relation $\gamma_sl \rho^{-2/3} / L =
\alpha$, where $\alpha$ is the Turnbull coefficient \cite{Turnbull} which has a 
roughly constant value of about 0.5 for elemental metals \cite{Hoyt2}.  The estimate 
of the undercooling required for a 1 nm premelted film is thus 
$\Delta T / T_M \sim 0.005$.  The results are consistent with the experimental 
studies of Balluffi and co-workers \cite{expt_list3} who estimated a lower bound of 
$T$=0.999 T$_M$ for the temperature where boundary widths of a few nm could be 
observed experimentally.

JJH acknowledges financial support from a Natural Sciences and Engineering
Research Council (NSERC)
of Canada Discovery grant.  Work at UC Davis and Northeastern was supported
by the US Department
of Energy (DOE), Office of Basic Energy Sciences, under contracts
DE-FG02-01ER45910 and -07ER46400,
respectively.  Sandia is a multiprogram laboratory operated by Sandia
Corporation, a Lockheed Martin
Company, for the DOE's National Nuclear Security Administration under
contract DE-AC04-94AL85000.
MA and SJ acknowledge helpful discussions with Dr. R. G. Hoagland.  All the
authors acknowledge
support from the DOE Computational Materials Science Network program.


\end{document}